\documentclass[dvips]{article}

\usepackage{icrctc07}

\title{Design Considerations for the Next Generation of Atmospheric
Imaging Cherenkov Telescopes}
\shorttitle{Design Considerations for IACT}

\authors{V.V. Bugaev$^{1}$, J.H. Buckley $^{1}$, H. Krawczynski$^{1}$ }
\shortauthors{V. Bugaev and et al.}
\afiliations{$^1$ Dept. of Physics, Washington Univ., St. Louis, MO,
63130 USA}
\email{vbugayov@rambler.ru}

\abstract{ We estimate the limiting angular resolution and detection area for an
array of 3 large-aperture Imaging Atmospheric Cherenkov Telescopes. We consider an idealized IACT system in
order to understand the limitations imposed by the intrinsic nature of
the atmospheric showers and geometry of the detector configuration. The idealization
includes the assumptions of a perfect optical system and the absence of the
night sky background with the goal of finding the optimum camera
geometry and array configuration independent of detailed assumptions
about the telescope design.

The showers are simulated using the ALTAI code for the altitude of 2700
m corresponding to one of possible future sites for a new
northern-hemisphere array. The optimal design depends on the target
energy range; for each energy we vary both the cell length (telescope
spacing) and the image
processing parameters in order to maximize the signal-to-noise ratio. We then present
the resulting values of the detection area and the angular resolution
for this energy dependent optimization.
We discuss the dependence of these quantities on the field of view of
the telescopes and pixel size of the camera.}

\begin{document}
\maketitle
%Begin the section.
\section{Introduction}

Future experiments based on Imaging Atmospheric Cherenkov Technique
are likely to benefit from advances in technology, but the technology
development strategy depends critically on performance and cost
tradeoffs for parameters such as pixel size, field of view and cell length.
A likely realization of the next generation IACT
observatory is a km$^2$ array of a large number of telescopes equipped with
$>100$ m$^2$ mirrors. The basic properties of such an array can be derived
partially from
the studies of a single cell of 3 telescopes. In this paper we 
determine the best performance of a system of 3 telescopes
with the aperture of 250 m$^2$. The idealizations
used here include:
\begin{itemize}
\item a perfect optical system,
\item ability to measure the exact coordinates of
individual photoelectrons in the focal plane of the camera,
\item zero level of night sky background (NSB),
\item field of view of a telescope that can be as large as $180^\circ$.
\end{itemize}

These idealizations allow us to study the limitations of the IACT
technique imposed by the nature of the atmospheric shower itself and
geometry of the telescopes rather
than by the specific details of the telescope hardware.

We determine the optimal
spacing between the telescopes for each target energy. 
For the optimization we chose the figure of merit to be the highest signal-to-noise ratio achieved for a
point-like $\gamma$-source, or
\begin{equation} S(E,l)=\frac{{dN_{\gamma}(E,
l)}/{dE}}{\sqrt{R_{\mathrm{CR}}(l)}}
\label{snr}
\end{equation}
where $R_{\mathrm{CR}}$ stands for the integral detection rate for the
cosmic-ray background and $dN_{\gamma}/dE$ is the differential
detection rate of photons from a Crab-like spectrum.
The function is closely related to the definition of the angular
resolution we use in the paper, which is the radius
around the source position in camera coordinates that provides the highest value for
the signal-to-noise ratio.

\section{Simulations}
For simulation of Cherenkov light produced both by $\gamma$-ray and CR
initiated showers we used the ALTAI code \cite{altai_code_nim}.  Only
one altitude (2700 m a.s.l.) has been considered so far, which
corresponds to the elevation of the site SPM Baja de Nord.
The simulated system consists of three telescopes forming a triangular cell.
While sky noise is not added to the simulated images we consider a
telescope as generating a trigger only if the image contains at least 100
p.e. This somewhat arbitrary value based on the experience of a
number of past experiments, and serves as a fiducial value for
comparison of different configurations; more careful efforts to determine trigger thresholds
typically generate rather unrealistically optimistic values that are not
realized.
The whole array is triggered if at least two of the
telescopes are triggered. The cell length ($l$) can vary withing the range
40 -- 400 m.
The photon-to-photoelectron conversion efficiency is assumed to be the
same as for the VERITAS telescopes. The overall photon-to-photoelectron
conversion factor is assumed to be equal to 0.1, which is a rather
conservative choice. The mirror size is equal to 250 m$^2$ corresponding
to an 18~m diameter dish. Only
protons were considered as the source of CR background in our simulations.
Both $\gamma$- and proton showers are generated with impact parameter radii up to 900 m.
$\gamma$-rays are vertical and parallel with energy in the range 3 --
2000 GeV, protons are isotropically distributed
within the range $0^\circ$ -- $5^\circ$ of zenith angles and sampled
according to an $E^{-2.7}$ spectrum \cite{antoni-2004-612} in the energy range 80 GeV -- 10 TeV. 
Among other effects, our simulations include the intrinsic Q for
background rejection. 
%Since the background rejection is so efficient for
%an array ($ > 1000:1 $)
We consider closely packed rectangular pixels.
Pixel size ($PS$) is defined here as the side length of a pixel.

\subsection{Event reconstruction}
For reconstruction of events we used Hillas moment analysis. The
arrival direction was then determined geometrically by intersection of the
major axes of the images in the camera coordinates. Coordinates for each
pair of telescopes were combined with a weight proportional to the square of the
sine of the angle between the major axes.
When calculating the image moments, we weighted each photoelectron
 depending on the proximity to the other
photoelectrons in the image. This weighting method improves the angular
resolution by $\sim 30\%$. For each photoelectron the radius $R_{25\%}$ of the
circle containing 25\% of other photoelectrons was determined. The
weight $w$ was calculated according to formula
$w=1/R^\alpha_{25\%}$, where 
$\alpha \simeq 2$.
%We perform a full three-dimensional reconstruction of the shower
%direction but do not include shape cuts (e.g. mean scaled width).

\section{Results}
\begin{figure}
\begin{center}
\rotatebox{-90}
{
\includegraphics [width=0.36\textwidth]{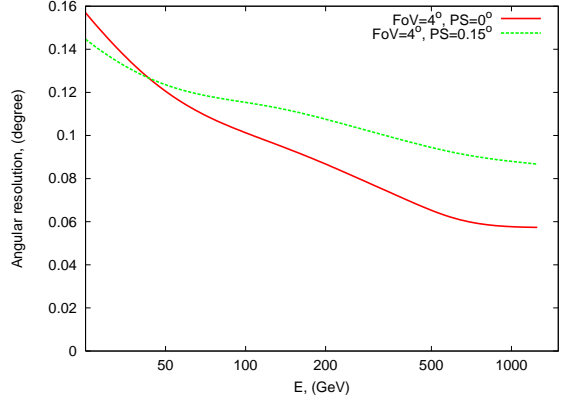} % theta_pixall.eps
}
\end{center}
\caption{Angular resolutions for cells of telescopes with $FoV=4^\circ$ and
different pixel sizes. 
Infinitely small pixels
($PS=0^\circ$) denotes
the case when coordinates of individual photoelectrons are assumed to be
known to arbitrary precision.
This plot shows the angular resolution for each
energy, using the optimum array configuration {\it for that energy}. }\label{theta_pixall}
\end{figure}

In Figure \ref{theta_pixall} we show the results on the angular
resolution for a cell of telescopes with $FoV=4^\circ$. 
As we will show later, $PS=0.15^\circ$ provides almost the same
signal-to-noise ratio as the case with the infinitely small
pixel size. 
By examining this figure, it can be recognized that below 50 GeV, the
cell spacing that optimizes the
sensitivity results in an angular resolution on the order of $0.15^\circ$.
For energies $\sim$
1~TeV this quantity is $\sim 0.06^\circ$ for $PS=0^\circ$ pixels, and is
somewhat worse ($\sim 0.09^\circ$) for $0.15^\circ$ pixels. Thus, there
is some improvement in angular resolution for finer pixelization at
moderate energy.
 
As a reminder, all the values on y-axes in the plots of this paper
correspond to the optimal value of the cell length at that energy.

\begin{figure}
\begin{center}
\rotatebox{-90}
{
\includegraphics [width=0.36\textwidth]{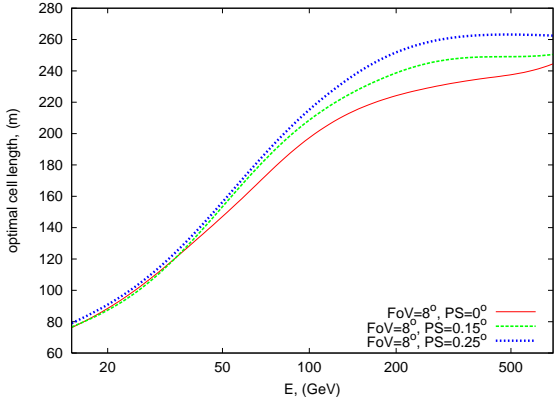} % celllen_pixall.eps
}
\end{center}
\caption{The dependence of the optimal cell length on energy for
$FoV=8^\circ$. All the plots in the paper take this dependence into
account.}\label{celllen_pixall}
\end{figure}

Next, we look at the optimal telescope spacing for wide FoV cameras.
The dependence of
the optimal cell length for $FoV=8^\circ$ and several different pixel
sizes is shown in
Figure \ref{celllen_pixall}.
For the larger values of the pixel size, the
optimal spacing between telescopes is also larger but only slightly.
This can be understood since signal-to-noise ratio increases with 
detection area more rapidly than improvement in angular resolution.
At some point
the degradation of the quality of image reconstruction dominates the
growth in the detection area, which limits the value for the optimal
cell length.

\begin{figure}
\begin{center}
\rotatebox{-90}
{
\includegraphics [width=0.36\textwidth]{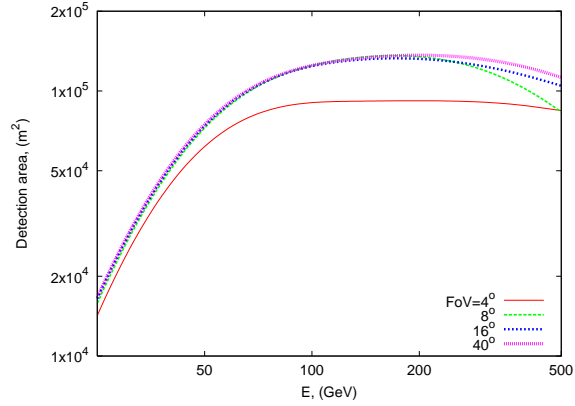} % areaprob_pix000.eps
}
\end{center}
\caption{The detection areas of the array after the cut on the arrival
direction in case of the infinitely
small pixel size and different $FoV$s.}\label{areaprob_pix000}
\end{figure}

\begin{figure}
\begin{center}
\rotatebox{-90}
{
\includegraphics [width=0.36\textwidth]{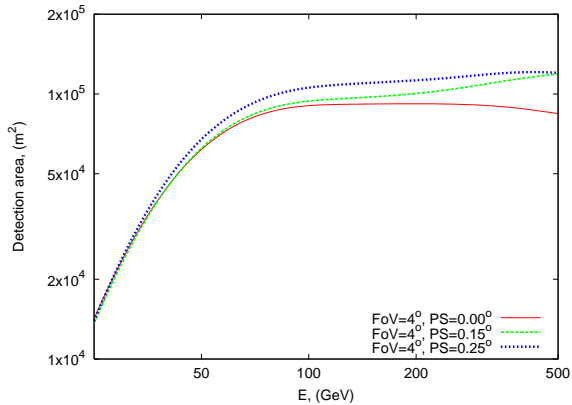} % areaprob_pixall_fovr2.eps
}
\end{center}
\caption{The detection areas of the array after the cut on the arrival
direction in case of $FoV=4^\circ$ and pixels of finite sizes.
}\label{areaprob_pixall_fovr2}
\end{figure}

The detection areas for an optimized array are shown in Figure
\ref{areaprob_pix000}, assuming several different values for the $FoV$
diameter. At each energy, we apply the optimal cut on the arrival direction
and assume that we have infinitely small pixels size. 
Between 4$^\circ$ and 8$^\circ$ $FoV$, the effective detection area (of
the optimized array configuration) increases by a factor of $\sim 1.6$,
but there is little difference for even larger fields of view. Thus, for
the assumed reconstruction algorithm, the optimum camera size lies
somewhere between 4$^\circ$ and 8$^\circ$. 
It is interesting to
note, that the optimal cut retains very nearly 50\% of the $\gamma$ events
everywhere in the energy range presented in this plot and in the plot
presented in Figure \ref{areaprob_pixall_fovr2}.

%\begin{figure}
%\begin{center}
%\rotatebox{-90}
%{
%\includegraphics [width=0.36\textwidth]{areaprob_pixall_fovr4.eps}
%}
%\end{center}
%\caption{The detection areas of the array after the cut on the arrival
%direction in case of $FoV=8^\circ$ and pixels of finite sizes.
%}\label{areaprob_pixall_fovr4}
%\end{figure}

Figure \ref{areaprob_pixall_fovr2} gives a feeling for
how the curves presented in Figure \ref{areaprob_pix000} are split
after also taking into account finite pixel size.
We observe that the splitting becomes smaller for the larger
values of $FoV$ adding weight to the conclusion that the optimal
value of $FoV$ is somewhere in the range $4^\circ$ -- $8^\circ$;
While signal-to-noise ratio does not change between $4^\circ$ and
$8^\circ$, the effective area increases by $\simeq 50\%$ for a point
source, and the solid angle for sky surveys increases by a factor of 4.

Having determined that $FoV$s larger than $8^\circ$ do not result in a
significant improvement of the point source sensitivity,
we plotted the signal-to-noise ratios for this $FoV$ and different pixel
sizes. Again a Crab spectrum was assumed for $\gamma$-rays. Choosing the
optimal pixel size is essentially the selection of the curve in this
plot that approaches the curve corresponding to $PS=0^\circ$. Given the
statistical errors corresponding to the deviations of the number of
$\gamma$-events used for signal-to-noise calculation, the curves with
$PS < 0.15^\circ$ appear to be indistinguishable. But, even given the
statistical uncertainties it is clear that reducing the pixel
size below $0.15^\circ$ will not
substantially improve the sensitivity of the array. The positions of the
peaks in the curves represent the energy threshold for the array of
three 18~m telescopes, which is $\sim 40$~GeV given the 100 ph.e.
lower limit of the image size. Taking into accout the error bars, the
thresholds are essentially the same for different pixel sizes, which
means that any
dependence of the energy threshold on the pixel size can be mainly
attributed to the impact of the NSB and may only be important very near
the threshold.

\begin{figure}
\begin{center}
\rotatebox{-90}
{
\includegraphics [width=0.36\textwidth]{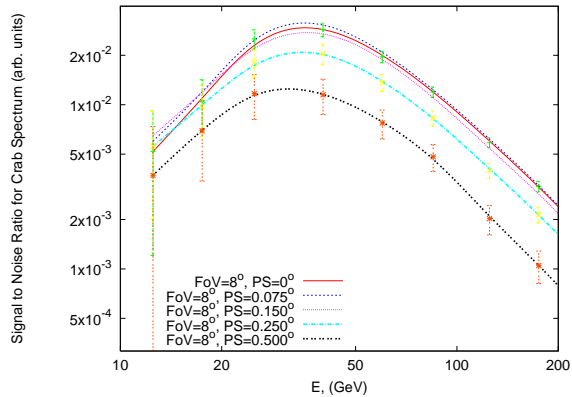} % q_pixall.eps
}
\end{center}
\caption{Comparison of signal-to-noise ratios for different pixel sizes.}\label{q_pixall}
\end{figure}

\section{Discussion and conclusion}
Telescopes with an $8^\circ$ field of view, achieve a better sensitivity
than telescopes with a field of view of $4^\circ$, however our
simulations show a smaller effect than previous studies
\cite{perez-2006-26}. Thus, our study shows that the main motivation for
building telescopes with a large $FoV$ is to observe extended sources and
to survey the TeV $\gamma$-ray sky more efficiently.

No significant differences in the signal-to-noise ratios are observed for
pixel size up to $0.15^\circ$ compared with the perfect
detector. This conclusion follows from the properties of $\gamma$-shower
itself and
intrinsic rejection of hadronic showers.
Optimization of angular resolution at high energies, as well as the
desire to reduce the NSB background in individual pixels may favor smaller pixels.
The values obtained here are optimal if night-sky noise is negligible.
For a future array this may indeed be the case, if the signal
integration
windows are further shortened, and sites with lower night sky background
noise are used. The results presented apply for an idealized
3-telescope cell. Work is in progress to extend our results to a km$^2$
array with the constraint on the total number of telescopes.
\bibliography{icrc0189}
\bibliographystyle{unsrt}

\end{document}